\documentclass[aps,prb,reprint,nolongbibliography,noeprint]{revtex4-1}
\usepackage{amsfonts,amssymb,amsmath,graphicx,color}
\usepackage[colorlinks=true,citecolor=blue,linkcolor=blue,urlcolor=blue]{hyperref}
\newcommand\cc{{\rm c.c.}}
\newcommand\hc{{\rm h.c.}}
\DeclareMathOperator\tr{tr}
\begin{document}
\title{Magnon gravitomagnetoelectric effect in noncentrosymmetric antiferromagnetic insulators}
\author{Atsuo Shitade}
\affiliation{Department of Physics, Kyoto University, Kyoto 606-8502, Japan}
\author{Youichi Yanase}
\affiliation{Department of Physics, Kyoto University, Kyoto 606-8502, Japan}
\date{\today}
\begin{abstract}
  We study the magnon contribution to the gravitomagnetoelectric (gravito-ME) effect,
  in which the magnetization is induced by a temperature gradient, in noncentrosymmetric antiferromagnetic insulators.
  This phenomenon is totally different from the ME effect, because the temperature gradient is coupled to magnons but an electric field is not.
  We derive a general formula of the gravito-ME susceptibility in terms of magnon wave functions and find that a difference in $g$ factors of magnetic ions is crucial.
  We also apply our formula to a specific model.
  Although the obtained gravito-ME susceptibility is small, we discuss several ways to enhance this phenomenon.
\end{abstract}
\maketitle
\section{Introduction} \label{sec:intro}
Spintronics exploits the spin degree of freedom of electrons, and it is an active research field in condensed matter physics.
Its central issues are generation, control, and detection of the spin and spin current without using a magnetic field.
Spin can be generated by an electric field in the Edelstein~\cite{Ivchenko1978,Ivchenko1989,Aronov1989,Edelstein1990233} and magnetoelectric (ME) effects~\cite{Fiebig_2005}.
These two phenomena are different;
the former is induced by the charge current in noncentrosymmetric metals,
while the latter is induced by the electric field when both the inversion and time-reversal symmetries are broken.
Another main subject is the spin Hall effect; the spin current flows perpendicular to the electric field~\cite{RevModPhys.87.1213}, which yields the spin accumulation at the boundaries.
The inverse spin Hall effect has been established as a method of detecting the spin current~\cite{1.2199473}.

Spincaloritronics, in which a temperature gradient plays a major role instead of the electric field,
has significantly developed in the past decade since the discovery of the spin Seebeck effect~\cite{Uchida2008,Jaworski2010}.
It enables us to convert waste heat into spin that carries information and improves existing thermoelectric devices.
The spin Nernst effect was theoretically proposed~\cite{PhysRevB.78.045302,LIU2010471,MA2010510,PhysRevB.94.035306,PhysRevB.94.205302,PhysRevB.98.081401}
and experimentally observed~\cite{Meyer2017,Shenge1701503,ncomms1400}.
Spin can be generated by the temperature gradient;
a heat analog of the Edelstein effect was already studied theoretically~\cite{WANG20101509,PhysRevB.87.245309,Xiao2016,PhysRevB.98.075307}.
Recently, we named a heat analog of the ME effect gravito-ME effect
in which the magnetization $M_a$ is induced by the temperature gradient $(-\partial_i T)$ as $\delta M_a = \beta^i_{\phantom{i} a} (-\partial_i T)$
and formulated the gravito-ME susceptibility $\beta^i_{\phantom{i} a}$~\cite{PhysRevB.99.024404,1812.11721}.
Although a similar effect was studied with use of the Kubo formula, the formula shows unphysical divergent susceptibility~\cite{PhysRevB.98.075307}.
We found that the correct gravito-ME susceptibility is obtained by subtracting the spin magnetic quadrupole moment (MQM) from the Kubo formula
and that it is related to the ME susceptibility by the Mott relation.

Spincaloritronics covers not only metals but also magnetic insulators whose low-energy physics is governed by magnons.
Since magnons are charge-neutral quasiparticles, the temperature gradient is an important driving force.
Indeed, various spincaloritronics phenomena by magnons have been elucidated. 
The spin Seebeck effect was observed in a ferrimagnetic insulator LaY$_2$Fe$_5$O$_{12}$~\cite{Uchida2010}.
Recently, the spin Nernst effect was theoretically proposed in ferromagnetic~\cite{PhysRevB.93.161106} and antiferromagnetic (AFM) insulators~\cite{PhysRevLett.117.217202,PhysRevLett.117.217203}
and soon later experimentally observed in MnPS$_3$~\cite{PhysRevB.96.134425}.
Apart from spincaloritronics, such a transverse motion of magnons was first observed by using the thermal Hall effect~\cite{Onose16072010,PhysRevB.85.134411},
which followed a theoretical proposal~\cite{PhysRevLett.104.066403}.
The importance of the magnetization correction was also pointed out~\cite{0022-3719-10-12-021,PhysRevB.55.2344,PhysRevLett.106.197202,PhysRevB.84.184406,PhysRevLett.107.236601}.

In this paper, we study the gravito-ME effect of magnons in noncentrosymmetric AFM insulators.
We find that it occurs when a unit cell contains multiple magnetic ions with different $g$ factors.
We emphasize that although the gravito-ME effect is an analog of the ME effect, these two phenomena may have essentially different origins.
The ME effect is attributed not to magnons but to the changes in the single-ion anisotropy, symmetric and antisymmetric exchange interactions, and $g$ factor by the electric field~\cite{Fiebig_2005}.
These ingredients may be affected by the temperature gradient as well, but we do not take them into account.
Therefore, in our setup, the electric field is not coupled to magnons but the temperature gradient is.

We clarify an important difference between the gravito-ME effect of electrons, which we studied previously~\cite{PhysRevB.99.024404}, and that of magnons, which we study here.
In both phenomena, a necessary condition is the presence of the interband matrix elements of the magnetization operator $M_a$.
Regarding the former, $M_a$ is proportional to the spin operator $S_a$ and can have the interband matrix elements.
Nonetheless, the gravito-ME susceptibility vanishes in any gapped electron system because of the Mott relation~\cite{PhysRevB.99.024404}.
Regarding the latter, $S_z$ cannot have the interband matrix elements, but $M_z$ can have in the above-mentioned situation.
In other words, the gravito-ME susceptibility vanishes when $g$ factors of magnetic ions are the same.
Each vanishing condition is not determined by symmetry.

This paper is organized as follows.
In Sec.~\ref{sec:model}, we introduce a model that exhibits the gravito-ME effect.
In Sec.~\ref{sec:formula}, we derive a formula of the gravito-ME susceptibility for general AFM insulators,
focusing on the case where the induced magnetization is parallel to the quantization axis.
In Sec.~\ref{sec:result}, we calculate the gravito-ME susceptibility for the model, but it turns out to be small.
Finally, in Sec.~\ref{sec:discussion} we propose several ways to enhance this phenomenon.

\section{Model} \label{sec:model}
\begin{figure}
  \centering
  \includegraphics[clip,width=0.48\textwidth]{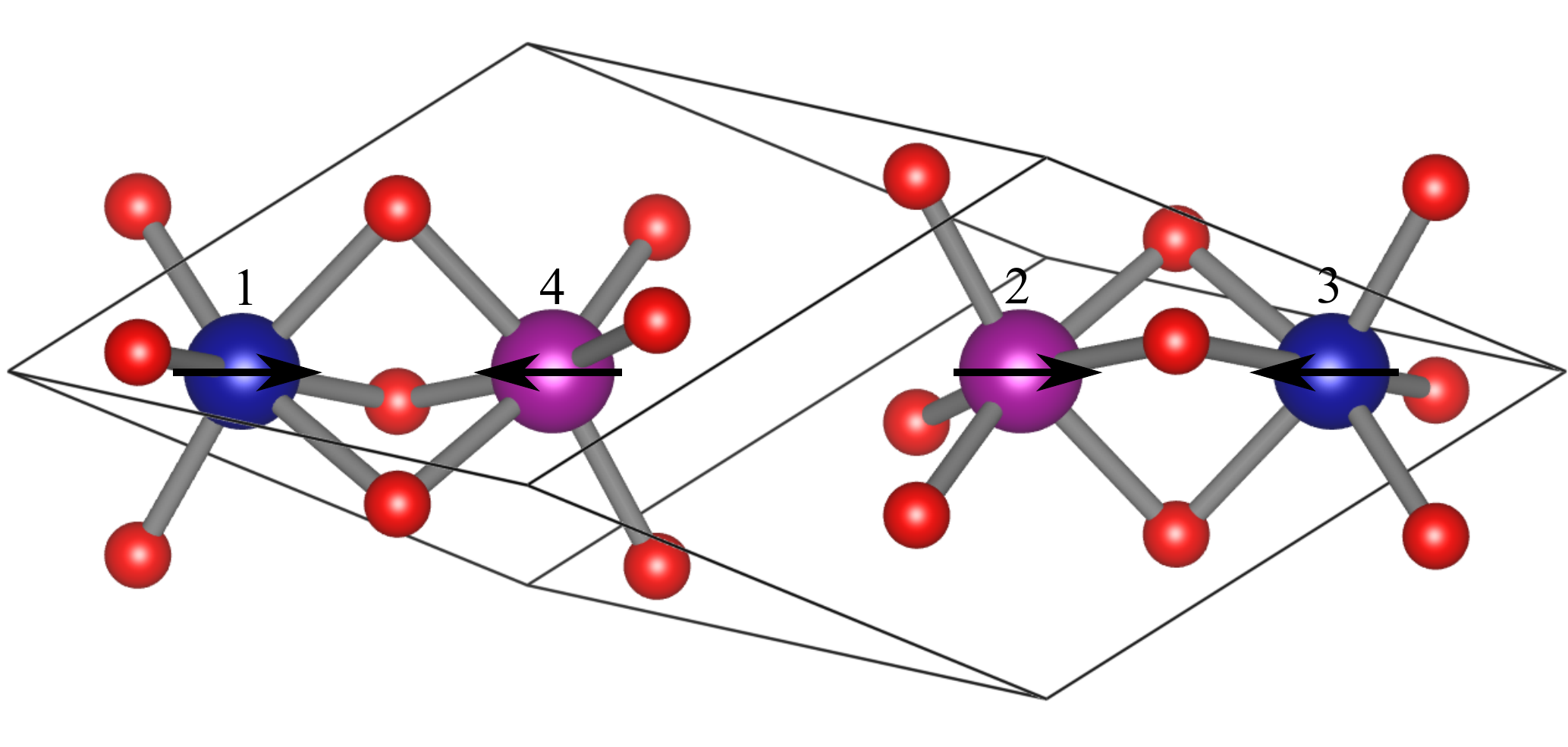}
  \caption{%
  Our model whose crystal and magnetic structures are the same as those of Cr$_2$O$_3$.
  Blue and purple circles represent magnetic ions $A$ and $B$, e.g., Cr$^{3+}$ and Fe$^{3+}$, and red circles represent O$^{2-}$.
  Black arrows illustrate the collinear AFM order.%
  } \label{fig:crfeo3}
\end{figure}
To begin with, let us introduce our model.
As shown in Fig.~\ref{fig:crfeo3}, the crystal and magnetic structures are the same as those of Cr$_2$O$_3$.
Among four magnetic ions in a unit cell, two ($\alpha = 1, 3$) are denoted by $A$ and the others ($\alpha = 2, 4$) are denoted by $B$.
Their spin sizes and $g$ factors are $S_A, S_B$, and $g_A = g + \delta g, g_B = g - \delta g$, respectively.
The spin Hamiltonian is given by~\cite{SAMUELSEN1969353,SAMUELSEN197013}
\begin{align}
  H
  = & \sum_i \Biggl[J_1 ({\vec S}_{{\vec r}_i, 1} \cdot {\vec S}_{{\vec r}_i, 4} + {\vec S}_{{\vec r}_i, 2} \cdot {\vec S}_{{\vec r}_i, 3}) \notag \\
  & + \sum_a ( J_{2A} {\vec S}_{{\vec r}_i, 1} \cdot {\vec S}_{{\vec r}_i + {\vec t}_a - {\vec c}, 3} + J_{2B} {\vec S}_{{\vec r}_i, 2} \cdot {\vec S}_{{\vec r}_i + {\vec t}_a, 4}) \notag \\
  & - H_A (S_{{\vec r}_i, 1 z} - S_{{\vec r}_i, 3 z}) - H_B (S_{{\vec r}_i, 2 z} - S_{{\vec r}_i, 4 z})\Biggr], \label{eq:crfeo31}
\end{align}
in which ${\vec S}_{{\vec r}_i, \alpha}$ is the spin operator of the $\alpha$th magnetic ion at the $i$th unit cell,
${\vec t}_a (a = 1, 2, 3)$ are primitive lattice vectors of the rhombohedral lattice,
and ${\vec c} = {\vec t}_1 + {\vec t}_2 + {\vec t}_3$.
The position of the $\alpha$th magnetic ion is ${\vec \rho}_{\alpha}$, which appears later.
$J_1, J_{2A}, J_{2B}$ are the exchange interactions.
$H_A, H_B$ are the effective anisotropy fields to constrain the ground state to the collinear AFM state,
which breaks the inversion symmetry and gives rise to the ME and gravito-ME effects.
The Dzyaloshinsky-Moriya interaction,
which is of the form $D ({\vec S}_{{\vec r}_i, 1} \times {\vec S}_{{\vec r}_i, 4} - {\vec S}_{{\vec r}_i, 2} \times {\vec S}_{{\vec r}_i, 3})^z$~\cite{DZYALOSHINSKY1958241,PhysRevLett.4.228,PhysRev.120.91},
is also allowed by symmetry, but it does not play an important role on the gravito-ME effect.

Low-energy physics of such a magnetic insulator is governed by magnons.
With use of the AFM Holstein-Primakoff transformation~\cite{PhysRev.58.1098}, the spin Hamiltonian Eq.~\eqref{eq:crfeo31} is approximated as
\begin{equation}
  H
  = \sum_{\vec k} a_{\vec k}^{\dag} H_{\vec k} a_{\vec k}, \label{eq:crfeo32}
\end{equation}
in which $a_{\vec k}^{\dag} \equiv [a_{1, {\vec k}}^{\dag}, a_{2, {\vec k}}^{\dag}, a_{3, -{\vec k}}, a_{4, -{\vec k}}]$ is a set of the magnon creation and annihilation operators,
and the magnon Hamiltonian $H_{\vec k}$ is given by
\begin{widetext} \begin{equation}
  H_{\vec k}
  = \begin{bmatrix}
    J_1 S_B + 3 J_{2A} S_A + H_A & 0 & 3 J_{2A} S_A \gamma_{\vec k} e^{-i {\vec k} \cdot {\vec \rho}_{24}} & J_1 \sqrt{S_A S_B} e^{i {\vec k} \cdot {\vec \rho}_{41}} \\
    0 & J_1 S_A + 3 J_{2B} S_B + H_B & J_1 \sqrt{S_A S_B} e^{i {\vec k} \cdot {\vec \rho}_{41}} & 3 J_{2B} S_B \gamma_{\vec k} e^{-i {\vec k} \cdot {\vec \rho}_{24}} \\
    3 J_{2A} S_A \gamma_{\vec k}^{\ast} e^{i {\vec k} \cdot {\vec \rho}_{24}} & J_1 \sqrt{S_A S_B} e^{-i {\vec k} \cdot {\vec \rho}_{41}} & J_1 S_B + 3 J_{2A} S_A + H_A & 0 \\
    J_1 \sqrt{S_A S_B} e^{-i {\vec k} \cdot {\vec \rho}_{41}} & 3 J_{2B} S_B \gamma_{\vec k}^{\ast} e^{i {\vec k} \cdot {\vec \rho}_{24}} & 0 & J_1 S_A + 3 J_{2B} S_B + H_B
  \end{bmatrix}. \label{eq:crfeo33}
\end{equation} \end{widetext}
Here, we have introduced
${\vec \rho}_{41} \equiv {\vec \rho}_4 - {\vec \rho}_1 = {\vec \rho}_3 - {\vec \rho}_2 = (2 v_4 - 1/2) {\vec c}$,
${\vec \rho}_{24} \equiv {\vec \rho}_2 - {\vec \rho}_4 = {\vec c} + {\vec \rho}_1 - {\vec \rho}_3 = (1 - 2 v_4) {\vec c}$,
and $\gamma_{\vec k} \equiv (e^{i {\vec k} \cdot {\vec t}_1} + e^{i {\vec k} \cdot {\vec t}_2} + e^{i {\vec k} \cdot {\vec t}_3})/3$.
See details in Appendix~\ref{app:holstein}.
$H_{\vec k}$ is diagonalized by a paraunitary matrix $P_{\vec k}$
that satisfies $P_{\vec k} \tau_3 P_{\vec k}^{\dag} = P_{\vec k}^{\dag} \tau_3 P_{\vec k} = \tau_3$,
in which $\tau_3$ is the third Pauli matrix for the particle-hole degree of freedom.
The eigenvalue problem to be solved is $\tau_3 H_{\vec k} | u_{n {\vec k}} \rangle = (\tau_3 E)_{n {\vec k}} | u_{n {\vec k}} \rangle$.
$\tau_3 H_{\vec k}$ is non-Hermitian but can be diaogonalized with the help of the Cholesky decomposition~\cite{COLPA1978327,PhysRevB.87.174427}.

\section{Magnon Gravito-ME Susceptibility} \label{sec:formula}
We focus on the $z$ component of the magnetization,
\begin{equation}
  M_z
  = -2 N_{\rm uc} g \mu_{\rm B}
  - \sum_{\vec k} a_{\vec k}^{\dag} (g \mu_{\rm B} \tau_3 + \delta g \mu_{\rm B} \sigma_3 \tau_3) a_{\vec k}, \label{eq:crfeo34}
\end{equation}
in response to the temperature gradient.
$N_{\rm uc}$ is the number of unit cells, $\mu_{\rm B}$ is the Bohr magneton,
and $\sigma_3$ is the third Pauli matrix for specifying the magnetic ions $A, B$.
In this way, we can separate the magnetization into the average part $-g \mu_{\rm B} \tau_3$
and the nonaverage part $\delta m_z$ that comes from the difference of $g$ factors.
In our setup, $\delta m_z = -\delta g \mu_{\rm B} \sigma_3 \tau_3$.
Below, we show that the latter part is crucial for the nonvanishing gravito-ME susceptibility.
The temperature gradient is introduced by Luttinger's gravitational potential $\phi_{\rm g}$ coupled to the Hamiltonian density~\cite{PhysRev.135.A1505}.
Hence, we calculate the correlation function of the magnetization and Hamiltonian,
\begin{align}
  \chi_{M_z H}^{\rm R}({\vec q}, \omega)
  = & -\frac{1}{N_{\rm uc}} \sum_{nm} \sum_{\vec k} \langle u_{n {\vec k}} | (-g \mu_{\rm B} \tau_3 + \delta m_z) \notag \\
  & \times | u_{m {\vec k} + {\vec q}} \rangle \langle u_{m {\vec k} + {\vec q}} | \tau_3 | u_{n {\vec k}} \rangle \notag \\
  & \times \frac{(\tau_3 E)_{n {\vec k}} + (\tau_3 E)_{m {\vec k} + {\vec q}}}{2} (\tau_3)_n (\tau_3)_m \notag \\
  & \times \frac{f((\tau_3 E)_{n {\vec k}}) - f((\tau_3 E)_{m {\vec k} + {\vec q}})}{\hbar \omega + (\tau_3 E)_{n {\vec k}} - (\tau_3 E)_{m {\vec k} + {\vec q}} + i \eta}, \label{eq:kubo1}
\end{align}
which characterizes the response $\langle \Delta M_z \rangle({\vec q}, \omega) = \chi_{M_z H}^{\rm R}({\vec q}, \omega) [-\phi_{\rm g}({\vec q}, \omega)]$.
Here, $f(z) = (e^{\beta z} - 1)^{-1}$ is the Bose distribution function, with $\beta = T^{-1}$ being the inverse of temperature,
and $\eta \rightarrow +0$ is the convergence factor.
See details in Appendix~\ref{app:bdg}.
By taking the limit of $\omega \rightarrow 0$ and then picking up the first order with respect to $q_i$,
we obtain the Kubo formula of the gravito-ME susceptibility:
\begin{align}
  T {\tilde \beta}^i_{\phantom{i} z}
  = & \lim_{\eta \rightarrow +0} \lim_{{\vec q} \rightarrow 0} \chi_{M_z H}^{\rm R}({\vec q}, 0)/i q_i \notag \\
  = & \frac{1}{N_{\rm uc}} \sum_n \sum_{\vec k}
  [\Omega^i_{\phantom{i} z n {\vec k}} (\tau_3 E)_{n {\vec k}} + m^i_{\phantom{i} z n {\vec k}}] f((\tau_3 E)_{n {\vec k}}) \notag \\
  & + \frac{1}{\eta N_{\rm uc}} \sum_n \sum_{\vec k} [-g \mu_{\rm B} (\tau_3)_n + \langle u_{n {\vec k}} | \delta m_z | u_{n {\vec k}} \rangle] \notag \\
  & \times (\tau_3)_n (\tau_3 E)_{n {\vec k}} [-f^{\prime}((\tau_3 E)_{n {\vec k}})] \partial_{k_i} (\tau_3 E)_{n {\vec k}}. \label{eq:kubo6}
\end{align}
Here, we have introduced
\begin{subequations} \begin{align}
  \Omega^i_{\phantom{i} z n {\vec k}}
  \equiv & i \sum_{m (\not= n)}
  \frac{\langle \partial_{k_i} u_{n {\vec k}} | \tau_3 | u_{m {\vec k}} \rangle \langle u_{m {\vec k}} | \delta m_z | u_{n {\vec k}} \rangle}{(\tau_3 E)_{n {\vec k}} - (\tau_3 E)_{m {\vec k}}}
  (\tau_3)_n (\tau_3)_m \notag \\
  & + \cc, \label{eq:kubo7a} \\
  m^i_{\phantom{i} z n {\vec k}}
  \equiv & -\frac{i}{2} \sum_{m (\not= n)}
  \langle \partial_{k_i} u_{n {\vec k}} | \tau_3 | u_{m {\vec k}} \rangle \langle u_{m {\vec k}} | \delta m_z | u_{n {\vec k}} \rangle
  (\tau_3)_n (\tau_3)_m \notag \\
  & + \cc \label{eq:kubo7b}
\end{align} \label{eq:kubo7}\end{subequations}
The second term in Eq.~\eqref{eq:kubo6} is divergent because we should take the limit of $\eta \rightarrow +0$ at the end of calculation.
Such an extrinsic contribution is identified as the heat analog of the Edelstein effect,
which was already studied in electron systems~\cite{WANG20101509,PhysRevB.87.245309,Xiao2016,PhysRevB.98.075307}.
In general, if we introduce disorder or interactions for magnons, $\eta$ may be nonzero, and it may remain finite.
In our model, however, the extrinsic contribution vanishes owing to the combined symmetry of the inversion and time-reversal transformations.

In order to obtain the correct gravito-ME susceptibility, we should subtract the spin MQM from the Kubo formula Eq.~\eqref{eq:kubo6},
because the gravitational potential perturbs not only the density matrix but also the magnetization density~\cite{PhysRevB.99.024404}.
The spin MQM is defined thermodynamically, namely, as the change in the grand potential by a magnetic-field gradient~\cite{PhysRevB.97.134423,1812.11721,PhysRevB.99.024404}.
We calculate another correlation function $\chi_{H M_z}^{\rm R}({\vec q}, \omega)$,
which characterizes the response $\langle \Delta H \rangle({\vec q}, \omega) = \chi_{H M_z}^{\rm R}({\vec q}, \omega) B^z({\vec q}, \omega)$,
and obtain the auxiliary spin MQM:
\begin{align}
  {\tilde M}^i_{\phantom{i} z}
  = & -\lim_{{\vec q} \rightarrow +0} \lim_{\eta \rightarrow +0} \chi_{H M_z}^{\rm R}({\vec q}, 0)/i q_i \notag \\
  & = \lim_{{\vec q} \rightarrow +0} \lim_{\eta \rightarrow +0} [\chi_{M_z H}^{\rm A}({\vec q}, 0)/i q_i]^{\ast} \notag \\
  = & \frac{1}{N_{\rm uc}} \sum_n \sum_{\vec k}
  \{\Omega^i_{\phantom{i} z n {\vec k}} (\tau_3 E)_{n {\vec k}} f((\tau_3 E)_{n {\vec k}}) \notag \\
  & + m^i_{\phantom{i} z n {\vec k}} [f((\tau_3 E)_{n {\vec k}}) + (\tau_3 E)_{n {\vec k}} f^{\prime}((\tau_3 E)_{n {\vec k}})]\}. \label{eq:mqm1}
\end{align}
Finally, we arrive at the spin MQM and gravito-ME susceptibility:
\begin{subequations} \begin{align}
  M^i_{\phantom{i} z}
  = & \frac{1}{N_{\rm uc}} \sum_n \sum_{\vec k} \notag \\
  & \times \left[-\Omega^i_{\phantom{i} z n {\vec k}} \int_{(\tau_3 E)_{n {\vec k}}}^{\infty} {\rm d} z f(z)
  + m^i_{\phantom{i} z n {\vec k}} f((\tau_3 E)_{n {\vec k}})\right], \label{eq:mqm2a} \\
  T \beta^i_{\phantom{i} z}
  = & T {\tilde \beta}^i_{\phantom{i} z} - M^i_{\phantom{i} z}
  = \frac{1}{N_{\rm uc}} \sum_n \sum_{\vec k} \Omega^i_{\phantom{i} z n {\vec k}} \notag \\
  & \times \left[(\tau_3 E)_{n {\vec k}} f((\tau_3 E)_{n {\vec k}}) + \int_{(\tau_3 E)_{n {\vec k}}}^{\infty} {\rm d} z f(z)\right] \notag \\
  & + \frac{1}{\eta N_{\rm uc}} \sum_n \sum_{\vec k} [-g \mu_{\rm B} (\tau_3)_n + \langle u_{n {\vec k}} | \delta m_z | u_{n {\vec k}} \rangle] \notag \\
  & \times (\tau_3)_n (\tau_3 E)_{n {\vec k}} [-f^{\prime}((\tau_3 E)_{n {\vec k}})] \partial_{k_i} (\tau_3 E)_{n {\vec k}}. \label{eq:mqm2b}
\end{align} \label{eq:mqm2}\end{subequations}
Equation~\eqref{eq:mqm2b} is our main result.
See details in Appendix~\ref{app:kubo}.
It is valid for general AFM insulators as far as magnon-magnon interactions can be neglected.

Now we explain why the average part $-g \mu_{\rm B} \tau_3$ does not contribute to the gravito-ME susceptibility Eq.~\eqref{eq:mqm2b}.
As demonstrated above, the gravito-ME susceptibility is essentially the correlation function $\chi_{H M_z}^{\rm R}({\vec q}, 0)/i q_i$.
Since $-g \mu_{\rm B} \tau_3$ is proportional to the magnon number operator,
we can interpret the external magnetic field and its gradient as the scalar potential and electric field in electronic systems,
although the particle statistics is different.
Therefore, we can interpret the average part of $\chi_{H M_z}^{\rm R}({\vec q}, 0)/i q_i$ as the thermodynamically defined charge polarization.
However, it is well known that the charge polarization is not appropriately defined in such a way; it vanishes even in ferroelectric states.
For the same reason, the average part of the gravito-ME susceptibility vanishes.
Note that the correct charge polarization is obtained by the charge current in an adiabatic deformation of the Hamiltonian~\cite{PhysRevB.47.1651,PhysRevB.48.4442,RevModPhys.66.899}.

\section{Results} \label{sec:result}
Let us apply the formula Eq.~\eqref{eq:mqm2b} to the model Eq.~\eqref{eq:crfeo33}.
We use $J_1 = 188.2~{\rm K}, J_{2A} = J_{2B} = 82.2~{\rm K}, H_A = H_B = 0.0495~{\rm K}$
obtained by the inelastic neutron-scattering experiment in Cr$_2$O$_3$~\cite{SAMUELSEN1969353,SAMUELSEN197013}.
The hexagonal lattice constants are $a = 5.01~{\rm \AA}, c = 13.55~{\rm \AA}$, and the position parameter is $v_4 = 0.348$.
Assuming magnetic $B$ sites replaced with irons,
we consider Cr$^{3+}$ and Fe$^{3+}$, whose spin sizes and $g$ factors are $S_A = 3/2, S_B = 5/2$ and $g_A = 1.97, g_B = 2$, respectively,
but neglect possible changes in the above parameters.
As mentioned above, the heat analog of the Edelstein effect is forbidden in this model, and the second term in Eq.~\eqref{eq:mqm2b} vanishes.

\begin{figure}
  \centering
  \includegraphics[clip,width=0.48\textwidth]{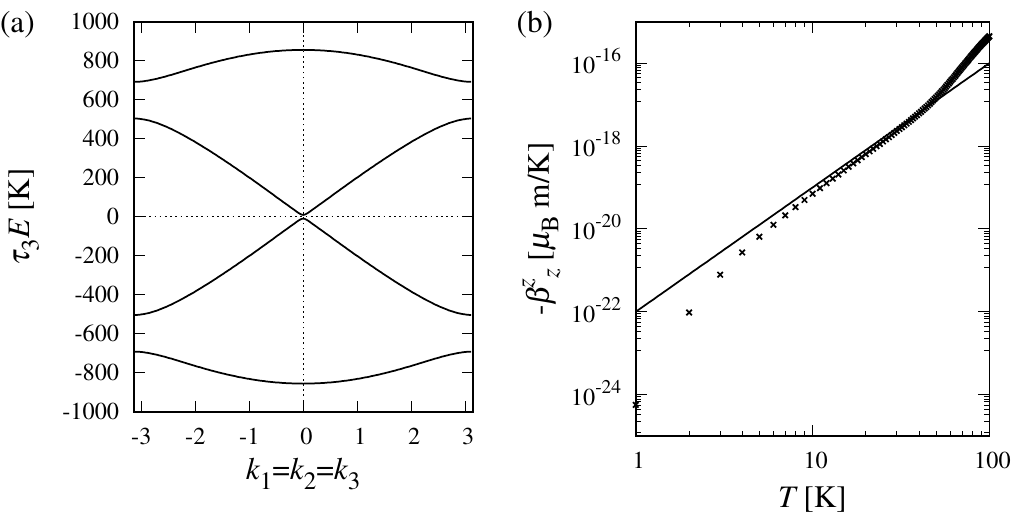}
  \caption{%
  (a) Magnon band structure along the $[111]$ direction.
  $k_a \equiv {\vec k} \cdot {\vec t}_a$.
  The gap at ${\vec k} = 0$ is $9.31~{\rm K}$.
  (b) Temperature dependence of the gravito-ME susceptibility.
  The black solid line represents $T^3$.%
  } \label{fig:afm}
\end{figure}
Figure~\ref{fig:afm}(a) shows the magnon band structure along the $[111]$ direction.
At ${\vec k} = 0$, the energy gap $E_{2 {\vec k} = 0} = 9.31~{\rm K}$ opens owing to the anisotropy fields.
Figure~\ref{fig:afm}(b) shows the temperature dependence of the gravito-ME susceptibility $\beta^z_{\phantom{z} z}$.
$\beta^z_{\phantom{z} z}$ shows an exponential decay below $E_{2 {\vec k} = 0}$,
while it is almost proportional to $T^3$ above $E_{2 {\vec k} = 0}$.
Note that the results above $50~{\rm K}$ are not reliable because the magnon-magnon interactions are no longer negligible.
At $T = 30~{\rm K}$, we obtain $\beta^z_{\phantom{z} z} = -2.31 \times 10^{-18} \mu_{\rm B}~{\rm m/K}$ per unit cell,
which means that the magnetization $M_z = -5.77 \times 10^{-16} \mu_{\rm B}$ per magnetic ion is induced
when the temperature gradient $(-\partial_z T) = 1~{\rm K/mm}$ is applied.
This value is much smaller than the current-induced magnetization estimated by the NMR experiment, which is of the order of $10^{-8} \mu_{\rm B}$~\cite{Furukawa2017}.

The temperature dependence of the gravito-ME susceptibility is understood as follows.
Around ${\vec k} = 0$, $E_{2 {\vec k}}$ is approximated as $E_{2 {\vec k}} = E_{2 {\vec k} = 0} + \hbar v k$,
and $\Omega^z_{\phantom{z} z 2 {\vec k}}$ is almost constant.
For $T > E_{2 {\vec k} = 0}$, we neglect the gap $E_{2 {\vec k} = 0}$ to evaluate $\beta^z_{\phantom{z} z}$ as
\begin{align}
  \beta^z_{\phantom{z} z}
  \simeq & \Omega^z_{\phantom{z} z 2 {\vec k} = 0} ({\vec t}_1 \cdot {\vec t}_2 \times {\vec t}_3) \notag \\
  & \times \int_0^{\infty} \frac{4 \pi k^2 {\rm d} k}{(2 \pi)^3} \left[\frac{\beta \hbar v k}{e^{\beta \hbar v k} - 1} - \ln (1 - e^{\beta \hbar v k})\right] \notag \\
  = & 8.66 \Omega^z_{\phantom{z} z 2 {\vec k} = 0} ({\vec t}_1 \cdot {\vec t}_2 \times {\vec t}_3)/2 \pi^2 (\beta \hbar v)^3, \label{eq:crfeo35}
\end{align}
which is proportional to $T^3$.
Although the interband effect may be enhanced at anticrossing points, whose energy scale is of the order of the exchange interactions,
magnons are not thermally excited to such high-energy states.
That is why the gravito-ME susceptibility obtained here is quite small.

\section{Discussion} \label{sec:discussion}
There are several ways to enhance the gravito-ME effect in AFM insulators.
First, for more complicated magnon bands, anticrossing points may appear at low energies, leading to the enhanced interband effect. 
Second, we can apply a larger temperature gradient than that in the above estimation,
although nonlinear effects that are not considered here may be important.
Third, for rare-earth magnetic ions, $g$ factors are given by Land\'e's $g$ factors and may be far from $2$.
We have considered transition-metal ions whose $g$ factors are slightly different from $2$ owing to crystalline fields and perturbative spin-orbit interactions,
leading to $\delta g = 0.03$.
If we choose Nd$^{3+}$ with $g_J = 8/11$, $\delta g$ is $7/11$.
Finally, we propose another mechanism of the gravito-ME effect.
In a spin-lattice-coupled system, acoustic phonons may be coupled to magnons~\cite{PhysRev.139.A450}.
As demonstrated above, the nonaverage part is crucial for the gravito-ME effect.
Since phonons do not carry spin, we can interpret that the $g$ factor of phonons is zero.
Even if the $g$ factors of magnetic ions are equal, the nonaverage part becomes nonzero in the full Hilbert space.
Furthermore, the spin-lattice coupling gives rise to anticrossing points whose energy scale is much smaller than the exchange interactions.
Thus, we expect that the gravito-ME effect is enhanced by the spin-lattice coupling, particularly near the corresponding temperature.

So far, we have focused only on the case where the induced magnetization is parallel to the quantization axis.
Since the symmetry requirement of the gravito-ME effect is the same as that of the ME effect,
the magnetization may be induced perpendicular to the quantization axis.
In the above model, $\beta^x_{\phantom{x} x} = \beta^y_{\phantom{y} y}$ and $\beta^x_{\phantom{x} y} = -\beta^y_{\phantom{y} x}$ are allowed by the $C_{3z}$ symmetry.
Such perpendicular components of the magnetization are expressed by linear combinations of the creation and annihilation operators and hence seem to vanish.
On the other hand, the perpendicular components may be nonzero when magnon Bose-Einstein condensation happens~\cite{PhysRevLett.84.5868}.
It is a future problem to formulate the gravito-ME susceptibility for the perpendicular components.
Extension to noncollinear magnetic insulators is also intriguing.

\section{Summary} \label{sec:summary}
To summarize, we have studied the gravito-ME effect in noncentrosymmetric AFM insulators, in which the magnetization is induced by a temperature gradient.
The induced magnetization may be parallel or perpendicular to the quantization axis, depending on lattice symmetries.
We have derived a general formula in the former case, which is expressed by magnon wave functions and is valid as long as magnon-magnon interactions are negligible.
We have found that the difference of $g$ factors of magnetic ions is crucial for the nonvanishing gravito-ME susceptibility.
As a representative, we have considered a model based on the first ME compound Cr$_2$O$_3$, in which two of four Cr$^{3+}$ ions are replaced with Fe$^{3+}$ ions.
The obtained gravito-ME susceptibility is small, and its experimental observation is challenging.
We expect that this phenomenon is enhanced in rare-earth compounds and spin-lattice-coupled systems.
\begin{widetext}
\begin{acknowledgments}
  We thank S. Murakami for informing us of Refs.~\cite{COLPA1978327,PhysRevB.87.174427} and Y. Shiomi for a comment from the experimental viewpoint.
  This work was supported by Grants-in-Aid for Scientific Research on Innovative Areas J-Physics (Grant No.~JP15H05884)
  and Topological Materials Science (Grant No.~JP18H04225) from the Japan Society for the Promotion of Science (JSPS),
  and by JSPS KAKENHI (Grants No.~JP15H05745, No.~JP18H01178, No.~JP18H05227, and No.~JP18K13508).
\end{acknowledgments}
\appendix
\section{AFM Holstein-Primakoff transformation} \label{app:holstein}
We employ the AFM Holstein-Primakoff transformation,
\begin{subequations} \begin{align}
  S_{{\vec r}_i, 1 z}
  = & S_A - n_{{\vec r}_i, 1},
  & S_{{\vec r}_i, 1 +}
  = & \sqrt{2 S_A - n_{{\vec r}_i, 1}} a_{{\vec r}_i, 1},
  & S_{{\vec r}_i, 1 -}
  = & a_{{\vec r}_i, 1}^{\dag} \sqrt{2 S_A - n_{{\vec r}_i, 1}}, \label{eq:holstein1a} \\
  S_{{\vec r}_i, 2 z}
  = & S_B - n_{{\vec r}_i, 2},
  & S_{{\vec r}_i, 2 +}
  = & \sqrt{2 S_B - n_{{\vec r}_i, 2}} a_{{\vec r}_i, 2},
  & S_{{\vec r}_i, 2 -}
  = & a_{{\vec r}_i, 2}^{\dag} \sqrt{2 S_B - n_{{\vec r}_i, 2}}, \label{eq:holstein1b} \\
  S_{{\vec r}_i, 3 z}
  = & -S_A + n_{{\vec r}_i, 3},
  & S_{{\vec r}_i, 3 +}
  = & a_{{\vec r}_i, 3}^{\dag} \sqrt{2 S_A - n_{{\vec r}_i, 3}},
  & S_{{\vec r}_i, 3 -}
  = & \sqrt{2 S_A - n_{{\vec r}_i, 3}} a_{{\vec r}_i, 3}, \label{eq:holstein1c} \\
  S_{{\vec r}_i, 4 z}
  = & -S_B + n_{{\vec r}_i, 4},
  & S_{{\vec r}_i, 4 +}
  = & a_{{\vec r}_i, 4}^{\dag} \sqrt{2 S_B - n_{{\vec r}_i, 4}},
  & S_{{\vec r}_i, 4 -}
  = & \sqrt{2 S_B - n_{{\vec r}_i, 4}} a_{{\vec r}_i, 4}, \label{eq:holstein1d}
\end{align} \label{eq:holstein1}\end{subequations}
in which $a_{{\vec r}_i, \alpha}, a_{{\vec r}_i, \alpha}^{\dag} (\alpha = 1, \dots, 4)$ are the annihilation and creation operators of a bosonic magnon,
and $n_{{\vec r}_i, \alpha} = a_{{\vec r}_i, \alpha}^{\dag} a_{{\vec r}_i, \alpha}$ is the number operator.
The spin Hamiltonian Eq.~\eqref{eq:crfeo31} is approximated as
\begin{align}
  H
  = & -N_{\rm uc} [2 J_1 S_A S_B + 3 (J_{2A} S_A^2 + J_{2B} S_B^2) + 2 (H_A S_A + H_B S_B)] \notag \\
  & + \sum_i \Biggl\{J_1 S_B (n_{{\vec r}_i, 1} + n_{{\vec r}_i, 3}) + J_1 S_A (n_{{\vec r}_i, 2} + n_{{\vec r}_i, 4})
  + J_1 \sqrt{S_A S_B} a_{{\vec r}_i, 1}^{\dag} a_{{\vec r}_i, 4}^{\dag} + J_1 \sqrt{S_A S_B} a_{{\vec r}_i, 2}^{\dag} a_{{\vec r}_i, 3}^{\dag} + \hc \notag \\
  & + \sum_a [J_{2A} S_A (n_{{\vec r}_i, 1} + n_{{\vec r}_i + {\vec t}_a - {\vec c}, 3}) + J_{2B} S_B (n_{{\vec r}_i, 2} + n_{{\vec r}_i + {\vec t}_a, 4})
  + J_{2A} S_A a_{{\vec r}_i, 1}^{\dag} a_{{\vec r}_i + {\vec t}_a - {\vec c}, 3}^{\dag} + J_{2B} S_B a_{{\vec r}_i, 2}^{\dag} a_{{\vec r}_i + {\vec t}_a, 4}^{\dag} + \hc] \notag \\
  & + H_A (n_{{\vec r}_i, 1} + n_{{\vec r}_i, 3}) + H_B (n_{{\vec r}_i, 2} + n_{{\vec r}_i, 4})\Biggr\}, \label{eq:holstein3}
\end{align}
in which $N_{\rm uc}$ is the number of unit cells, and hence the number of magnetic ions is $4 N_{\rm uc}$.
The first line represents the ground-state energy.
Here, we employ the Fourier transformation
\begin{subequations} \begin{align}
  a_{{\vec r}_i, \alpha}^{\dag}
  = & \frac{1}{\sqrt{N_{\rm uc}}} \sum_{\vec k} e^{-i {\vec k} \cdot ({\vec r}_i + {\vec \rho}_{\alpha})} a_{\alpha, {\vec k}}^{\dag}, \label{eq:holstein4a} \\
  a_{{\vec r}_i, \alpha}
  = & \frac{1}{\sqrt{N_{\rm uc}}} \sum_{\vec k} e^{i {\vec k} \cdot ({\vec r}_i + {\vec \rho}_{\alpha})} a_{\alpha, {\vec k}}. \label{eq:holstein4b}
\end{align} \label{eq:holstein4}\end{subequations}
The Hamiltonian Eq.~\eqref{eq:holstein3} turns into
\begin{align}
  H
  = & -N_{\rm uc} [2 J_1 S_A S_B + 3 (J_{2A} S_A^2 + J_{2B} S_B^2) + 2 (H_A S_A + H_B S_B)] \notag \\
  & + \sum_{\vec k} \Biggl\{(J_1 S_B + 3 J_{2A} S_A + H_A) (n_{1, {\vec k}} + n_{3, {\vec k}}) + (J_1 S_A + 3 J_{2B} S_B + H_B) (n_{2, {\vec k}} + n_{4, {\vec k}}) \notag \\
  & + J_1 \sqrt{S_A S_B} e^{i {\vec k} \cdot ({\vec \rho}_4 - {\vec \rho}_1)} a_{1, {\vec k}}^{\dag} a_{4, -{\vec k}}^{\dag}
  + J_1 \sqrt{S_A S_B} e^{i {\vec k} \cdot ({\vec \rho}_3 - {\vec \rho}_2)} a_{2, {\vec k}}^{\dag} a_{3, -{\vec k}}^{\dag} + \hc \notag \\
  & + \sum_a [J_{2A} S_A e^{i {\vec k} \cdot ({\vec t}_a - {\vec c} + {\vec \rho}_3 - {\vec \rho}_1)} a_{1, {\vec k}}^{\dag} a_{3, -{\vec k}}^{\dag}
  + J_{2B} S_B e^{i {\vec k} \cdot ({\vec t}_a + {\vec \rho}_4 - {\vec \rho}_2)} a_{2, {\vec k}}^{\dag} a_{4, -{\vec k}}^{\dag} + \hc]\Biggr\} \notag \\
  = & -N_{\rm uc} \{J_1 [S_A (S_B + 1) + S_B (S_A + 1)] + 3 [J_{2A} S_A (S_A + 1) + J_{2B} S_B (S_B + 1)] \notag \\
  & + 2 [H_A (S_A + 1/2) + H_B (S_B + 1/2)]\}
  + \sum_{\vec k} a_{\vec k}^{\dag} H_{\vec k} a_{\vec k}. \label{eq:holstein5}
\end{align}
Thus, we obtain the magnon Hamiltonian Eq.~\eqref{eq:crfeo33}.
Also, the $z$ component of the magnetization is expressed in terms of magnons as
\begin{align}
  M_z
  = & \sum_i [g_A \mu_{\rm B} (S_{{\vec r}_i 1 z} + S_{{\vec r}_i 3 z}) + g_B \mu_{\rm B} (S_{{\vec r}_i 2 z} + S_{{\vec r}_i 4 z})]
  = -\sum_i [g_A \mu_{\rm B} (n_{{\vec r}_i 1} - n_{{\vec r}_i 3}) + g_B \mu_{\rm B} (n_{{\vec r}_i 2} - n_{{\vec r}_i 4})] \notag \\
  = & -\sum_{\vec k} [g_A \mu_{\rm B} (n_{1 {\vec k}} - n_{3 {\vec k}}) + g_B \mu_{\rm B} (n_{2 {\vec k}} - n_{4 {\vec k}})]
  = -2 N_{\rm uc} g \mu_{\rm B} - \sum_{\vec k} a_{\vec k}^{\dag} (g \mu_{\rm B} \tau_3 + \delta g \mu_{\rm B} \sigma_3 \tau_3) a_{\vec k}, \label{eq:holstein6}
\end{align}
which is Eq.~\eqref{eq:crfeo34}.

\section{Kubo formula for a bosonic Bogoliubov-de Gennes Hamiltonian} \label{app:bdg}
The Kubo formula enables us to calculate any linear response $\langle \Delta Y \rangle(\omega) = \chi_{YX}^{\rm R}(\omega) F(\omega)$,
in which $X$ is conjugate to an external field $F$,
namely, the perturbation Hamiltonian is given by $H_1(t) = -X F(t)$.
$\chi_{YX}^{\rm R}(\omega)$ is given by
\begin{equation}
  \chi_{YX}^{\rm R}(\omega)
  = \frac{1}{i \hbar} \int_0^{\infty} {\rm d} t e^{i (\hbar \omega + i \eta) t/\hbar} \tr [\rho [X, Y(t)]]. \label{eq:bdg1}
\end{equation}
$\rho$ is the density matrix for a Hamiltonian $H$, and $Y(t) \equiv e^{i H t/\hbar} Y e^{-i H t/\hbar}$.
The trace in Eq.~\eqref{eq:bdg1} is expanded with respect to the eigenstates $| \psi_n \rangle$ of a bosonic Bogoliubov-de Gennes Hamiltonian as
\begin{align}
  \tr [\rho [X, Y(t)]]
  = & \sum_{n_1 \dots n_4} \langle \psi_{n_3} | Y | \psi_{n_4} \rangle \langle \psi_{n_1} | X | \psi_{n_2} \rangle
  \tr \{\rho [(\alpha^{\dag})_{n_1} (\alpha)_{n_2}, [\alpha^{\dag}(t)]_{n_3} [\alpha(t)]_{n_4}]\} \notag \\
  = & \sum_{n_1 \dots n_4} \langle \psi_{n_3} | Y | \psi_{n_4} \rangle \langle \psi_{n_1} | X | \psi_{n_2} \rangle
  e^{i [(\tau_3 E)_{n_3} - (\tau_3 E)_{n_4}] t/\hbar}
  \tr \{\rho [(\alpha^{\dag})_{n_1} [(\alpha)_{n_2}, (\alpha^{\dag})_{n_3}] (\alpha)_{n_4} \notag \\
  & + [(\alpha^{\dag})_{n_1}, (\alpha^{\dag})_{n_3}] (\alpha)_{n_2} (\alpha)_{n_4}
  + (\alpha^{\dag})_{n_3} (\alpha^{\dag})_{n_1} [(\alpha)_{n_2}, (\alpha)_{n_4}]
  + (\alpha^{\dag})_{n_3} [(\alpha^{\dag})_{n_1}, (\alpha)_{n_4}] (\alpha)_{n_2}\} \notag \\
  = & \sum_{n_1 \dots n_4} \langle \psi_{n_3} | Y | \psi_{n_4} \rangle \langle \psi_{n_1} | X | \psi_{n_2} \rangle
  e^{i [(\tau_3 E)_{n_3} - (\tau_3 E)_{n_4}] t/\hbar}
  \tr \{\rho [(\tau_3)_{n_2 n_3} (\alpha^{\dag})_{n_1} (\alpha)_{n_4} - (\tau_3)_{n_4 n_1} (\alpha^{\dag})_{n_3} (\alpha_{n_2})]\} \notag \\
  = & \sum_{n_1 \dots n_4} \langle \psi_{n_3} | Y | \psi_{n_4} \rangle \langle \psi_{n_1} | X | \psi_{n_2} \rangle
  e^{i [(\tau_3 E)_{n_3} - (\tau_3 E)_{n_4}] t/\hbar}
  (\tau_3)_{n_2 n_3} (\tau_3)_{n_4 n_1} [f((\tau_3 E)_{n_4}) - f((\tau_3 E)_{n_4})]. \label{eq:bdg2}
\end{align}
Here, we have used the commutation relations of bosons, i.e., $[(\alpha)_{n_2}, (\alpha^{\dag})_{n_3}] = (\tau_3)_{n_2 n_3}$,
and $\tr [\rho (\alpha^{\dag})_{n_1} (\alpha)_{n_4}] = f((\tau_3 E)_{n_4}) (\tau_3)_{n_1 n_4}$.
Thus, the Kubo formula is rewritten by
\begin{equation}
  \chi_{YX}^{\rm R}(\omega)
  = -\sum_{nm} \langle \psi_n | Y | \psi_m \rangle \langle \psi_m | X | \psi_n \rangle
  (\tau_3)_n (\tau_3)_m
  \frac{f((\tau_3 E)_n) - f((\tau_3 E)_m)}{\hbar \omega + (\tau_3 E)_n - (\tau_3 E)_m + i \eta}. \label{eq:bdg3}
\end{equation}
Equation~\eqref{eq:kubo1} is obtained by setting $X = H, F = -\phi_{\rm g}, Y = M_z$.

\section{Evaluation of the correlation function} \label{app:kubo}
Let us evaluate Eq.~\eqref{eq:kubo1} in the limit of $\omega \rightarrow 0$.
The intraband contribution $n = m$ is
\begin{align}
  \chi_{M_z H}^{\rm R (I)}({\vec q}, 0)
  = & -\frac{1}{N_{\rm uc}} \sum_{n} \sum_{\vec k}
  \langle u_{n {\vec k}} | (-g \mu_{\rm B} \tau_3 + \delta m_z) | u_{n {\vec k} + {\vec q}} \rangle \langle u_{n {\vec k} + {\vec q}} | \tau_3 | u_{n {\vec k}} \rangle
  \frac{(\tau_3 E)_{n {\vec k}} + (\tau_3 E)_{n {\vec k} + {\vec q}}}{2} \notag \\
  & \times \frac{f((\tau_3 E)_{n {\vec k}}) - f((\tau_3 E)_{n {\vec k} + {\vec q}})}{(\tau_3 E)_{n {\vec k}} - (\tau_3 E)_{n {\vec k} + {\vec q}} + i \eta}. \label{eq:kubo2}
\end{align}
Up to the first order with respect to $q_i$, we obtain
\begin{equation}
  \chi_{M_z H}^{\rm R (I)}({\vec q}, 0)
  = \frac{i q_i}{\eta N_{\rm uc}} \sum_n \sum_{\vec k} [-g \mu_{\rm B} (\tau_3)_n + \langle u_{n {\vec k}} | \delta m_z | u_{n {\vec k}} \rangle] (\tau_3)_n
  (\tau_3 E)_{n {\vec k}} [-f^{\prime}((\tau_3 E)_{n {\vec k}})] \partial_{k_i} (\tau_3 E)_{n {\vec k}}. \label{eq:kubo3}
\end{equation}

The interband contribution $n \not= m$ is
\begin{align}
  \chi_{M_z H}^{\rm R (II)}({\vec q}, 0)
  = & -\frac{1}{N_{\rm uc}} \sum_{n \not= m} \sum_{\vec k}
  \langle u_{n {\vec k}} | (-g \mu_{\rm B} \tau_3 + \delta m_z) | u_{m {\vec k} + {\vec q}} \rangle \langle u_{m {\vec k} + {\vec q}} | \tau_3 | u_{n {\vec k}} \rangle
  \frac{(\tau_3 E)_{n {\vec k}} + (\tau_3 E)_{m {\vec k} + {\vec q}}}{2} (\tau_3)_n (\tau_3)_m \notag \\
  & \times \frac{f((\tau_3 E)_{n {\vec k}}) - f((\tau_3 E)_{m {\vec k} + {\vec q}})}{(\tau_3 E)_{n {\vec k}} - (\tau_3 E)_{m {\vec k} + {\vec q}}}. \label{eq:kubo4}
\end{align}
Now we can safely take the limit of $\eta \rightarrow +0$.
Up to the first order with respect to $q_i$, we find
\begin{align}
  \chi_{M_z H}^{\rm R (II)}({\vec q}, 0)
  = & -\frac{q_i}{N_{\rm uc}} \sum_{n \not= m} \sum_{\vec k}
  \langle u_{n {\vec k}} | \delta m_z | u_{m {\vec k}} \rangle \langle \partial_{k_i} u_{m {\vec k}} | \tau_3 | u_{n {\vec k}} \rangle
  \frac{(\tau_3 E)_{n {\vec k}} + (\tau_3 E)_{m {\vec k}}}{2} (\tau_3)_n (\tau_3)_m \notag \\
  & \times \frac{f((\tau_3 E)_{n {\vec k}}) - f((\tau_3 E)_{m {\vec k}})}{(\tau_3 E)_{n {\vec k}} - (\tau_3 E)_{m {\vec k}}} \notag \\
  = & -\frac{q_i}{N_{\rm uc}} \sum_{n \not= m} \sum_{\vec k}
  \frac{\langle \partial_{k_i} u_{n {\vec k}} | \tau_3 | u_{m {\vec k}} \rangle \langle u_{m {\vec k}} | \delta m_z | u_{n {\vec k}} \rangle - \cc}{(\tau_3 E)_{n {\vec k}} - (\tau_3 E)_{m {\vec k}}}
  \frac{(\tau_3 E)_{n {\vec k}} + (\tau_3 E)_{m {\vec k}}}{2} (\tau_3)_n (\tau_3)_m  f((\tau_3 E)_{n {\vec k}}) \notag \\
  = & \frac{i q_i}{N_{\rm uc}} \sum_n \sum_{\vec k}
  [\Omega^i_{\phantom{i} z n {\vec k}} (\tau_3 E)_{n {\vec k}} + m^i_{\phantom{i} z n {\vec k}}] f((\tau_3 E)_{n {\vec k}}). \label{eq:kubo5}
\end{align}
In the first line, the average part $-g \mu_{\rm B} \tau_3$ vanishes owing to $\langle u_{n {\vec k}} | (-g \mu_{\rm B} \tau_3) | u_{m {\vec k}} \rangle = 0$ for $n \not= m$.
From Eq.~\eqref{eq:kubo5}, we obtain Eq.~\eqref{eq:kubo6}.
$\Omega^i_{\phantom{i} z n {\vec k}}, m^i_{\phantom{i} z n {\vec k}}$ are defined in Eq.~\eqref{eq:kubo7} and can be rewritten as
\begin{subequations} \begin{align}
  \Omega^i_{\phantom{i} z n {\vec k}}
  = & i \sum_{m (\not= n)} \frac{\langle u_{n {\vec k}} | \partial_{k_i} H_{\vec k} | u_{m {\vec k}} \rangle \langle u_{m {\vec k}} | \delta m_z | u_{n {\vec k}} \rangle}{[(\tau_3 E)_{n {\vec k}} - (\tau_3 E)_{m {\vec k}}]^2}
  (\tau_3)_n (\tau_3)_m + \cc, \label{eq:kubo8a} \\
  m^i_{\phantom{i} z n {\vec k}}
  = & -\frac{i}{2} \sum_{m (\not= n)} \frac{\langle u_{n {\vec k}} | \partial_{k_i} H_{\vec k} | u_{m {\vec k}} \rangle \langle u_{m {\vec k}} | \delta m_z | u_{n {\vec k}} \rangle}{(\tau_3 E)_{n {\vec k}} - (\tau_3 E)_{m {\vec k}}}
  (\tau_3)_n (\tau_3)_m + \cc \notag \\
  = & -\frac{i}{2} \langle \partial_{k_i} u_{n {\vec k}} | [(\tau_3)_n - \tau_3 | u_{n {\vec k}} \rangle \langle u_{n {\vec k}} |] \delta m_z | u_{n {\vec k}} \rangle + \cc, \label{eq:kubo8b}
\end{align} \label{eq:kubo8}\end{subequations}
by using $\langle u_{n {\vec k}} | \partial_{k_i} H_{\vec k} | u_{m {\vec k}} \rangle = \partial_{k_i} (\tau_3 E)_{n {\vec k}} (\tau_3)_{nm} + [(\tau_3 E)_{n {\vec k}} - (\tau_3 E)_{m {\vec k}}] \langle \partial_{k_i} u_{n {\vec k}} | \tau_3 | u_{m {\vec k}} \rangle$.

In the intraband contribution Eq.~\eqref{eq:kubo2}, let us take the limit of $\eta \rightarrow +0$ and then pick up the first order with respect to $q_i$.
In this case, we obtain
\begin{align}
  \chi_{M_z H}^{\rm R (I)}({\vec q}, 0)
  = & -\frac{q_i}{N_{\rm uc}} \sum_n \sum_{\vec k}
  \biggl\{\langle u_{n {\vec k}} | (-g \mu_{\rm B} \tau_3 + \delta m_z) | \partial_{k_i} u_{n {\vec k}} \rangle (\tau_3)_n (\tau_3 E)_{n {\vec k}} f^{\prime}((\tau_3 E)_{n {\vec k}}) \notag \\
  & + [-g \mu_{\rm B} (\tau_3)_n + \langle u_{n {\vec k}} | \delta m_z | u_{n {\vec k}} \rangle] \langle \partial_{k_i} u_{n {\vec k}} | \tau_3 | u_{n {\vec k}} \rangle (\tau_3 E)_{n {\vec k}} f^{\prime}((\tau_3 E)_{n {\vec k}}) \notag \\
  & + [-g \mu_{\rm B} (\tau_3)_n + \langle u_{n {\vec k}} | \delta m_z | u_{n {\vec k}} \rangle] (\tau_3)_n \frac{1}{2} \partial_{k_i} (\tau_3 E)_{n {\vec k}} f^{\prime}((\tau_3 E)_{n {\vec k}}) \notag \\
  & + [-g \mu_{\rm B} (\tau_3)_n + \langle u_{n {\vec k}} | \delta m_z | u_{n {\vec k}} \rangle] (\tau_3)_n (\tau_3 E)_{n {\vec k}} \frac{1}{2} f^{\prime \prime}((\tau_3 E)_{n {\vec k}}) \partial_{k_i} (\tau_3 E)_{n {\vec k}}\biggr\} \notag \\
  = & \frac{q_i}{2 N_{\rm uc}} \sum_n \sum_{\vec k}
  (\{\langle \partial_{k_i} u_{n {\vec k}} | [(\tau_3)_n - \tau_3 | u_{n {\vec k}} \rangle \langle u_{n {\vec k}} |] \delta m_z | u_{n {\vec k}} \rangle - \cc\}
  (\tau_3 E)_{n {\vec k}} f^{\prime}((\tau_3 E)_{n {\vec k}}) \notag \\
  & - \partial_{k_i} \{[-g \mu_{\rm B} (\tau_3)_n + \langle u_{n {\vec k}} | \delta m_z | u_{n {\vec k}} \rangle] (\tau_3)_n (\tau_3 E)_{n {\vec k}} f^{\prime}((\tau_3 E)_{n {\vec k}})\}) \notag \\
  = & \frac{i q_i}{N_{\rm uc}} \sum_n \sum_{\vec k} m^i_{\phantom{i} z n {\vec k}} (\tau_3 E)_{n {\vec k}} f^{\prime}((\tau_3 E)_{n {\vec k}}). \label{eq:kubo9}
\end{align}
Here, we have dropped the total derivative with respect to $k_i$.
The average part $-g \mu_{\rm B} \tau_3$ vanishes again
because $\langle u_{n {\vec k}} | (-g \mu_{\rm B} \tau_3) | u_{n {\vec k}} \rangle = -g \mu_{\rm B} (\tau_3)_n$ is independent of $k_i$.
From Eqs.~\eqref{eq:kubo5} and \eqref{eq:kubo9}, we obtain Eq.~\eqref{eq:mqm1}.
Equation~\eqref{eq:mqm2a} is obtained by solving $\partial (\beta M^i_{\phantom{i} z})/\partial \beta = {\tilde M}^i_{\phantom{i} z}$.
\end{widetext}
%
\end{document}